\documentclass[12pt]{article}
\pdfoutput=1

\usepackage{amssymb,latexsym}

\usepackage{graphicx}

\def\dalemb#1#2{{\vbox{\hrule height .#2pt
        \hbox{\vrule width.#2pt height#1pt \kern#1pt
                \vrule width.#2pt}
        \hrule height.#2pt}}}

\let\a=\alpha \let\b=\beta \let\g=\gamma \let\d=\delta \let\e=\epsilon
  \let\th=\theta  
\let\l=\lambda \let\m=\mu  \let\x=\xi  %\let\r=\rho

        \let\Th=\Theta 
\let\X=\Xi  \let\S=\Sigma  \let\Y=\Psi
 
\let\la=\label  
  
\def\nn{\nonumber} \def\bd{\begin{document}} \def\ed{\end{document}}
\def\ds{\documentstyle} \let\fr=\frac \let\bl=\bigl \let\br=\bigr
\let\Br=\Bigr \let\Bl=\Bigl
\let\bm=\bibitem
\let\na=\nabla
\def\tU{{\widetilde U}}
\let\pa=\partial \let\ov=\overline
\def\ie{{\it i.e.\ }}
\newcommand{\be}{\begin{equation}}
\newcommand{\ee}{\end{equation}}
\def\ba{\begin{array}}
\def\ea{\end{array}}
\def\ft#1#2{{\textstyle{{\scriptstyle #1}\over {\scriptstyle #2}}}}
\def\fft#1#2{{#1 \over #2}}
\def\F#1#2{{ F_{#1}^{(#2)} }}
\def\cF#1#2{{ {\cal F}_{#1}^{(#2)} }}

\def\R{{\bf R}}
\def\sst#1{{\scriptscriptstyle #1}}
\def\oneone{\rlap 1\mkern4mu{\rm l}}
\def\e7{E_{7(+7)}}
\def\td{\tilde}
\def\wtd{\widetilde}
\def\im{{\rm i}}
\def\bog{Bogomol'nyi\ }
\newcommand{\ho}[1]{$\, ^{#1}$}
\newcommand{\hoch}[1]{$\, ^{#1}$}
\newcommand{\bea}{\begin{eqnarray}}
\newcommand{\eea}{\end{eqnarray}}
\newcommand{\ra}{\rightarrow}
\newcommand{\lra}{\longrightarrow}
\newcommand{\Lra}{\Leftrightarrow}
\newcommand{\ap}{\alpha^\prime}
\newcommand{\bp}{\tilde \beta^\prime}
\newcommand{\cB}{{\cal B}}
\newcommand{\cO}{{\cal O}}
\newcommand{\vecx}{\vec{x}}
\newcommand{\vecy}{\vec{y}}
\newcommand{\vecp}{\vec{p}}
\newcommand{\vecq}{\vec{q}}
\newcommand{\tr}{{\rm tr} }
\newcommand{\Tr}{{\rm Tr} }
\newcommand{\NP}{Nucl. Phys. }

\newcommand{\cL}{{\cal L}}
\newcommand{\cA}{{\cal A}}
\newcommand{\cD}{{\cal D}}

\def\sst#1{{\scriptscriptstyle #1}}
\def\0{{\sst{(0)}}}
\def\1{{\sst{(1)}}}
\def\2{{\sst{(2)}}}
\def\3{{\sst{(3)}}}
\def\4{{\sst{(4)}}}
\def\5{{\sst{(5)}}}
\def\6{{\sst{(6)}}}
\def\7{{\sst{(7)}}}
\def\8{{\sst{(8)}}}
\def\ve{\varepsilon}
\def\vf{\varphi}
\def\F{\Phi}
\def\wg{\wedge}

\newcommand{\tamphys}{\it A.I. Akhiezer Institute for Theoretical Physics,\\
NSC ``Kharkov Institute of Physics and Technology",\\
Kharkov 61108, Ukraine
}

\newcommand{\auth}{AUTHORS}

\def\thb{\bar{\theta}}
\def\Thb{\bar{\Theta}}
\def\barp{\bar{p}}
\def\barq{\bar{q}}
\def\barc{\bar{c}}
\def\bard{\bar{d}}
\def\e{\epsilon}

\def \bi{\bibitem}
\def \la {\label}

\def \l {\lambda}
\def\foot{\footnote}
\def \tl  {{\tilde \l}}
\def \sql {{\sqrt \l}}
\def \adss {$AdS_5 \times S^5$\ }
\newcommand{\rf}[1]{(\ref{#1})}
\def \ov {\over}

\def\th{\theta}
\def\Th{\Theta}
\def\vth{\vartheta}
\def\btheta{{\bar\theta}}
\def\ttheta{{{\tilde\theta}}}
\def\bttheta{{{\bar\ttheta}}}
\def\vth{\vartheta}

\def\ra{\rightarrow}
\def\N{{\cal N}}
\def\F{{\cal F}}
\def\uM{\underline{M}}
\def\uN{\underline{N}}
\def\uP{\underline{P}}
\def\ua{\underline{a}}
\def\ub{\underline{b}}
\def\uc{\underline{c}}
\def\ud{\underline{d}}
\def\ue{\underline{e}}
\def\uf{\underline{f}}
\def\ual{\underline{\alpha}}
\def\ube{\underline{\beta}}
\def\um{\underline{m}}
\def\un{\underline{n}}
\def\umu{\underline{\mu}}
\def\unu{\underline{\nu}}
\def\cc{\circ}
\def\eqv{\equiv}

\def\ni{\noindent}

\def\Ep{E^{{}^{(+)}}}
\def\Em{E^{{}^{(-)}}}

\def\Mp{M^{{}^{(+)}}}
\def\Mm{M^{{}^{(-)}}}

\def \ha{{1\ov 2}}

\def\r{\rho}

\def\Y{{\rm Y}}
\def\X{{\rm X}}
\def\tY{\tilde{\rm Y}}
\def\tX{\tilde{\rm X}}
\def\dY{\dot{\rm Y}}
\def\dX{\dot{\rm X}}

\def \J {\mathcal{J}}
\def \del {\partial}

\def\dF{\dot{F}}
\def\dG{\dot{G}}
\def\df{\dot{f}}
\def \E {{\cal E}}
\def \S {{\cal S}}
\def \J {{\cal J}}

\def\ms{\mathcal{S}}
\def\mj{\mathcal{J}}
\def\soj{\fr{\ms}{\mj}}
\def \R {{\bf R}}
\def \om {\omega}
\def \bE {\bar E}
\def \x {{\cal X}}

\def \bi{\bibitem}
\def \la {\label}

\def \l {\lambda}
\def\foot{\footnote}
\def \tl  {{\tilde \l}}
\def \sql {{\sqrt \l}}
\def \adss {$AdS_5 \times S^5$\ }
\def \ov {\over}

\def \varpi {{\rm w}}

\def\thb{\bar{\theta}}
\def\Thb{\bar{\Theta}}
\def\mb{\bar{\m}}
\def\ab{\bar{\a}}
\def\zb{\bar{z}}
\def\psib{\bar{\psi}}
\def\barp{\bar{p}}
\def\barq{\bar{q}}
\def\barc{\bar{c}}
\def\bard{\bar{d}}
\def\e{\epsilon}
\def\wb{\bar{w}}
\def\lb{\bar{\l}}
\def\Jb{\bar{J}}
\def\Nb{\bar{N}}
\def\pab{\bar{\pa}}

\def\At{\tilde{A}}
\def\Bt{\tilde{B}}
\def\Ct{\tilde{C}}
\def\Dt{\tilde{D}}
\def\Et{\tilde{E}}
\def\Ft{\tilde{F}}
\def\Gt{\tilde{G}}
\def\Mt{\tilde{M}}
\def\at{\tilde{a}}
\def\bt{\tilde{b}}
\def\ct{\tilde{c}}
\def\dt{\tilde{d}}
\def\et{\tilde{e}}
\def\ft{\tilde{f}}
\def\gt{\tilde{g}}
\def\mt{\tilde{\mu}}
\def\nt{\tilde{\nu}}

\def\ola{\overleftarrow}
\def\ora{\overrightarrow}
\def\alt{\tilde{\a}}

\def\eh{\hat{e}}
\def\eph{\hat{\e}}
\def\ph{\hat{p}}
\def\alh{\hat{\a}}
\def\beh{\hat{\b}}
\def\gah{\hat{\g}}
\def\muh{\hat{\m}}
\def\thh{\hat{\th}}
\def\dh{\hat{d}}
\def\deh{\hat{\d}}
\def\wh{\hat{w}}
\def\lah{\hat{\l}}
\def\Ch{\hat{C}}
\def\Omh{\hat{\Omega}}

\def\ps{\rlap{\, /}\;\,p }
\def\ks{\rlap{\, /}\;\,k }

\def\gym{g_{YM}}

\def\adot{\dot{a}}
\def\bdot{\dot{b}}
\def\bpa{\bar{\pa}}

\def\pr{\prime}

\begin{document}
\overfullrule=0pt
\parskip=2pt
\parindent=12pt
\headheight=0in \headsep=0in \topmargin=0in
\oddsidemargin=0in

\vspace{ -3cm}
\thispagestyle{empty}
%\vspace{1cm}
%\begin{flushright}
%Preprint DFPD 01/TH/\\
%hep-th/
%\end{flushright}

\title{Analytical approach to phase transitions in rotating and non-rotating 2D holographic superconductors}

\author{A.J. Nurmagambetov\\
\\
{\it A.I. Akhiezer Institute for Theoretical Physics,}\\
{\it NSC ``Kharkov Institute of Physics and Technology",}\\
{\it Kharkov 61108, Ukraine}\\
{\tt ajn@kipt.kharkov.ua}}

\date{\phantom{July 2011}}

%\tamphys

\maketitle

\abstract{Analytical calculations of phase transitions in AdS3 Maxwell-scalar system, modeling a holographic superconductor, are performed in the probe limit of BTZ black hole background. Estimated values of the phase transition critical temperature and of the scalar boundary operator are compared with that obtained in numerical simulations. We discuss extensions of the model to topologically massive QED3 and to inclusion of external magnetic field. The latter becomes possible through the Barnett-London and the Lense-Thirring effects, but the former does not lead to any viable modifications.
The standard setup of a holographic superconductor is extended to a rotating 2D holographic superconductor, and the main characteristics of the phase transition are calculated in the probe limit and in the small angular momentum approximation. We evaluate the feedback of the superconductor rotation on the phase transition and find the set of parameters, which lower the critical temperature in compare to the non-rotating case, or make it slightly higher.}

 \vspace{0.1cm}

\setcounter{equation}{0}
\setcounter{footnote}{0}
\setcounter{section}{0}

\newpage

\tableofcontents

\section{Introduction}

Superconductivity is one of the most fascinating phenomena in Condensed Matter Physics (CMP), which have great impact on applied physics and technology. Further progress in searching for new superconducting materials requires the theoretical ground to predict, quantitatively and qualitatively, the behavior of such systems. Conventional superconductors are well described by BCS theory \cite{Bardeen:1957mv}, \cite{Bogoliubov:1958}, the main ingredients of which are the phase transition near the critical point and the  electron Cooper pairs formation. The latter is responsible for forming an energy gap, explaining, for instance, the dependence of anomalous heat capacity on temperature in superconductors. The main effective coupling in BCS theory is the electron-phonon coupling, so BCS theory is one of the best examples of theories with weak coupling constant.

Further development of experimental studies has led to the discovery of high-temperature superconductivity (HTSC) in the so-called cuprates \cite{Bednorz:1986tc}, with different, in compare to BCS theory, mechanism of the Cooper pairs condensate creation. It was realized that models of HTSC must be formulated as theories in the strong coupling constant regime \cite{AndersonBook}, that puts serious restrictions as on the theory development, as well as on its predictive ability. Therefore, it is important to figure out new mechanisms of phase transitions \cite{AndersonBook}, \cite{SachdevBook} and of the energy gap formation to describe superconductivity in new types of superconductors \cite{Bednorz:1986tc}, \cite{Chen:2008}.

One of the ways to describe theories in the strong coupling constant regime is the AdS/CFT correspondence  \cite{Maldacena:1997re}, \cite{Gubser:1998bc}, \cite{Witten:1998qj}, \cite{Aharony:1999ti}. Applying the AdS/CFT to QCD allowed to achieve the predicted by QCD power law behavior of hadronic amplitudes \cite{Matveev:1973ra}, \cite{Brodsky:1973kr} at high energies within string theory on AdS \cite{Polchinski:2001tt}, and got back the stringy description to QCD \cite{Brower:2002er}, \cite{BoschiFilho:2002zs}, \cite{Andreev:2002aw}. The basic principle of the AdS/CFT correspondence is the description of a strong coupled Conformal Field Theory living on the AdS boundary in terms of the dual string theory in curved space and in the small coupling constant regime. In such a limit string theory is effectively described by AdS supergravity, i.e. by General Relativity with matter fields propagating in the AdS bulk.

Applying the AdS/CFT to Superconductivity was initiated in \cite{Hartnoll:2008vx}, \cite{Hartnoll:2008kx}. Further developments in this area have resulted in transforming the original AdS/CFT proposal to AdS/CMP correspondence (see e.g. \cite{GGdualBook11} for recent reviews), and in significant progress in the description of CMP models with strong coupling constant in terms of the dual gravity with matter fields \cite{Hartnoll:2009sz}, \cite{Herzog:2009xv}, \cite{Horowitz:2010gk}, and in the presence of Black Holes (BH). The choice of matter fields is responsible for simulated properties of a holographic superconductor, and for triggering, at the level of the boundary CFT, the phase transition and the energy gap formation. Such an approach was called the holographic superconductivity, and it is strongly based on the properties of AdS BHs.

Inclusion of AdS black holes into the holographic superconductivity setup plays a dual role. On the one hand, a black hole provides non-zero temperature for the boundary CFT. On the other hand, AdS black holes can form a scalar hair, which condenses at the boundary. For neutral AdS black holes the neutral scalar hair makes theory unstable \cite{Hertog:2006rr}, however a charged AdS black hole supports charged scalar hair with large enough charge \cite{Gubser:2008px}. The latter observation puts the holographic superconductivity on the firm ground, and makes possible to calculate main characteristics of superconductors.

In the original version of the holographic superconductor model \cite{Hartnoll:2008vx}, \cite{Hartnoll:2008kx}, fields in the bulk consist of Maxwell field interacting with a charged scalar, and this configuration puts in the background of $\mathrm{AdS_{d+1}}$ neutral black hole. It corresponds to the probe limit, in which any backreaction of matter fields on the metric is neglected. Following the AdS/CFT dictionary, a vector field in the bulk corresponds to a current on the boundary CFT side, and a scalar field corresponds to the boundary CFT operator, whose expectation value is defined by the field asymptote near the boundary. The mass of the charged scalar is bounded from below by the value of the Breitenlohner-Freedman mass $m_{BF}$ \cite{Breitenlohner:1982bm} that guarantees the renormalizable solution at the boundary and in the bulk.

Choosing the appropriate ansatz for the charged scalar and for the abelian vector field \cite{Hartnoll:2008vx}, \cite{Hartnoll:2008kx}, the system of equations of motion in the background of $\mathrm{AdS_4}$ neutral black hole and in the probe limit was numerically solved. It was obtained the value of the critical temperature, below which the boundary scalar field operator forms the condensate, and the functional dependence of condensation on the boundary CFT temperature. The obtained dependence turned out to be quite similar to that obtained in BCS theory. Numerical calculations of the DC conductivity also revealed the gap formation \cite{Hartnoll:2008vx}, \cite{Hartnoll:2008kx}, so the correspondence of the dual bulk theory to qualitative description of the holographic superconductor was established and verified.

The next natural step towards constructing and analyzing new holographic superconductor models is the development of analytical methods of solving for systems of nonlinear interacting differential equations, governing the qualitative behavior and defining quantitative characteristics of such systems. Analytical approaches are useful as for deeper understanding physical processes behind the numerical simulations, as well as for controlling the estimated in the numerical studies principle parameters of the system. Furthermore, in the case of numerical studies of holographic superconductivity, it is a delicate point to introduce a cut-off parameter at temperatures close to zero. Having the relevant analytical approach allows one to control the choice of the cut-off parameter and the correctness of numerical estimations.

Analytical methods in holographic superconductivity were mostly  developed and applied for 3D and 4D models of holographic superconductors \cite{Gregory09}, \cite{Herzog:2010vz}, \cite{Bellon:2010xs}, \cite{Siopsis:2010uq}, \cite{Siopsis:2010pi}, 
\cite{Chen:2011en}, \cite{Ge:2011cw}. However, the case of 2D holographic superconductivity, based on the $\mathrm{AdS_3/CFT_2}$ correspondence interesting on its own (see, for instance, 
\cite{Carlip:2005zn} on QFT side, and \cite{Arutuynov:2008} on CMP side), still remains uncover within the analytical approach, though has numerically studied in \cite{Ren10}, \cite{Liu:2011fy}.

The aim of this paper is twofold. First, we fill the gap in applying the analytical methods to 2D superconductors in the probe limit, and compare the estimated value of the critical temperature and the temperature dependence of the boundary scalar operator  expectation value with that obtained within the numerical simulations, see Section 2 below. Second, we extend our simple model of a 2D holographic superconductor, modifying either fields in the bulk, or the background in which they propagate. A simple and natural modification of 3D abelian vector field in $\mathrm{AdS_3}$ to topologically massive vector field \cite{Deser:1982vy}, \cite{Deser:1981wh} was suggested in \cite{Lashkari:2010ak}. It is straightforward to check that non-triviality of 3D Chern-Simons term $\mathcal{L}_{CS}=\theta/2 (A\wg F)$, generating the mass of 3D vector field $A$, is provided by magnetic degrees of freedom in the vector field ansatz. However, such a modification is not suitable for studying the phase transition in 2D CFT, since the known solutions of $\mathrm{AdS_3}$ magnetic black holes are horizonless. The way to overcome this obstacle and to engage external magnetic field in a game is discussed in Section 3. To this end, the AdS black hole background has to be changed to a rotational black hole. 

Studies of rotating black holes in context of holographic superconductivity were initiated in \cite{Sonner:2009fk}. We exploit ideas of \cite{Sonner:2009fk} and construct a 2D rotating holographic superconductor in the probe limit and in the small angular momentum approximation. Section 4 contains the general setup of the problem, details and results of calculations are collected in Section 5. Summing up the results and concluding remarks may be found in the last section. Symbols $``="$ and $``\approx"$ are used throughout the paper to distinguish exact and approximative equalities.

\section{2D holographic superconductor in the probe limit}

In the probe limit fields are supposed to propagate in AdS black hole background, without backreaction on the metric field. The metric of the neutral $\mathrm{AdS_3}$ black hole in Poincar\'e coordinates is defined by
\begin{equation}\label{AdS3BH}
ds^2=\frac{L^2}{z^2}\left(-f(z)dt^2+dx^2+\frac{dz^2}{f(z)}\right),\qquad f(z)=1-\frac{z^2}{z^2_H}.
\end{equation} 
Here $L$ is a characteristic length of the AdS space, $(t,x,z)$ are the coordinates of $\mathrm{AdS_3}$. $(t,x)$ parameterize the boundary of $\mathrm{AdS_3}$ located at $z=0$. In the pure AdS, i.e. when $f(z)=1$, $z$ varies from infinity to zero. Metric \rf{AdS3BH} corresponds to the neutral BH solution to the Einstein equation
\begin{equation}\label{EinsAdS}
R_{mn}-\fr12 g_{mn}(R+\fr2{L^2})=0.
\end{equation}
The black hole horizon is located at $z_H$, and the Hawking temperature is
\begin{equation}\label{TBH}
T=\fr1{4\pi}|f^\prime(z)|_{\vert_{z=z_H}}	=\fr1{2\pi z_H}.
\end{equation}

Let us consider the following configuration of fields
\begin{equation}\label{acAdS}
S=-\int\, d^3 x \, \sqrt{-g} \left(\fr14 F_{mn}F^{mn}+(\pa_m-iA_m)\Psi (\pa^m+iA^m)\Psi^*+m^2 \Psi\Psi^*\right)
\end{equation}
in the background \rf{AdS3BH} and at the temperature \rf{TBH}. Varying fields in \rf{acAdS} results in the set of equations of motion
\bea\label{EOM3}
\fr1{\sqrt{-g}}D_m(\sqrt{-g}D^m\Psi)-m^2\Psi=0,\nn \\
\pa_n(\sqrt{-g}F^{nm})+\sqrt{-g}i(\Psi\overline{D^m}\Psi^*-\Psi^*D^m \Psi)=0,
\eea
with $D\Psi\equiv (\pa_m-iA_m)\Psi$, $\overline{D^m}\Psi^*=(D\Psi)^*$, $F_{mn}=2\pa_{[m}A_{n]}$. We are going to solve \rf{EOM3} with the following ansatz
\bea
\Psi=\psi(z),\qquad A=\phi(z)dt
\label{ans1}
\eea 
and to demonstrate the occurrence of the phase transition due to the charged complex scalar field $\Psi$, which forms a condensate at the boundary of $\mathrm{AdS_3}$. Recall that $\Psi$-condensate  is an analog of the Cooper pairs on BCS theory side.

Part of the gauge field equation of motion with ``radial" free index $z$ leads to the constant phase of the complex scalar field, so in what follows we will treat, without loss of generality, $\psi(z)$ as a real function of $z$. Taking the latter into account, \rf{EOM3} reduces to the system of differential equations
\bea
\psi^{\prime \prime}+\left(\fr{f^\prime}{f}-\fr1{z}\right)\psi^\pr+\left(\fr{\phi^2}{f^2}-\fr{m^2L^2}{z^2 f}\right)\psi=0, \nn\\
\phi^{\prime \prime}+\fr{\phi^\prime}{z}-\fr{L^2}{z^2}\fr{2\psi^2}{f}\phi=0,
\label{EOMzprobe}
\eea
where prime denotes the derivative with respect to $z$. The mass of the scalar field is restricted by the renormalizability (at the boundary and in the bulk) of the solutions. Here we study the case $m^2L^2=-1$ corresponding to the Breitenlohner-Freedman (BF) bound \cite{Breitenlohner:1982bm}, that guarantees the positivity of the charged massive scalar field energy.

The system of equations \rf{EOMzprobe} can be solved numerically \cite{Ren10} (see also \cite{Liu:2011fy} for the numerical solution to the complete system with backreaction), but we are aimed at finding the analytical solution, in the spirit of \cite{Gregory09}. To this end, one needs to specify boundary conditions for fields at the boundary and at the horizon. At $z=z_H$ we have $f(z_H)=0$, so we set $\phi(z_H)=0$ to have the finite norm of the vector field everywhere in the bulk \cite{Hartnoll:2008vx}, \cite{Hartnoll:2008kx}. Evaluating the equation for $\psi$ near the horizon leads to $\psi^\prime(z_H)=\psi(z_H)/2z_H$, hence the first set of the boundary conditions (BCs) is
\bea
\phi(z_H)=0,\qquad \psi^\prime(z_H)=\psi(z_H)/2z_H .
\label{BCzH}
\eea

Another set of the BCs comes from evaluating \rf{EOMzprobe} at $z=0$. Here we have $\psi=\psi^\1 z\ln z+\psi^\2 z$ and $\phi=\mu \ln z-\rho$. Following the AdS/CFT dictionary, the asymptotic expansion of any bulk field near the boundary of $\mathrm{AdS_{d+1}}$
\bea
\Theta(z)=\mathcal{A}z^{\Delta_-}(1+\dots)+\mathcal{B}z^{\Delta_+}(1+\dots)
\label{asymp}
\eea
includes the source $\mathcal{A}$ to the corresponding boundary operator $\mathcal{O}$. The expectation value of the operator is given by $\mathcal{B}=\langle \mathcal{O}\rangle$. In \rf{asymp} $\Delta_{\pm}$  are the characteristic exponents whose values are fixed to be $\Delta(\Delta-d)=m^2L^2$ for a scalar field, and $\Delta(\Delta-d+2)=m^2L^2$ for a vector field. Eq. \rf{asymp} may be adopted to the $d\rightarrow 2$ limit \cite{Ren10}, leading to the expressions for $\psi(0)$ and $\phi(0)$. Since we expect that the dual to the scalar field operator condenses at the boundary without being sourced, we can choose either $\langle \mathcal{O}\rangle=\psi^\1$, or $\langle \mathcal{O}\rangle=\psi^\2$, with setting the other coefficient $\psi^\2$ (or $\psi^\1$) to zero. However, $\psi^\2=0$ leads to the boundary theory which is not conformal in the usual sense \cite{Marolf:2006nd}, so we choose $\psi^\1=0$ throughout the paper. 

Coefficients in the BC of $\phi$ at $z\rightarrow 0$ are the chemical potential $\mu$ and the charge density $\rho$ (see e.g. \cite{Hartnoll:2007ai}). Such an identification comes from considering the boundary asymptotic expansion of a vector field in $\mathrm{AdS_{d+1}}$, which for a massless vector field is
\bea
A_t (z\rightarrow 0)=\mathcal{A}z^{(d-2)}(1+\dots)+\mathcal{B}(1+\dots)
\label{Aasymp}
\eea
(other components of the vector field are zero in \rf{ans1}). It turns out that the $z$-dependent part of the vector potential has a fast falloff, hence it does not correspond to a background field in the dual theory. It fixes the electric charge density of the state in the field theory, while the finite part of $A_t(z)$ at the boundary may be considered as a chemical potential for the electric charge in CFT. This quantities are related to each other by the requirement of vanishing $A_t(z)$ at the horizon. By use of $z^\epsilon=1+\epsilon \ln z+\dots$ for small $\epsilon$, supplying with $\mu\rightarrow -\mu/\epsilon$ \cite{Ren10}, it is easy to check that in the $d\rightarrow 2$ limit $A_t(0)=\mu \ln z-\rho$ after discarding the divergent under $\epsilon=(d-2) \rightarrow 0$ terms. Therefore, the second set of the BCs at $z\rightarrow 0$ is
\bea
\phi=\mu \ln z-\rho,\qquad \psi=\psi^\2 z .
\label{BCz0}
\eea

After fixing the BCs at the AdS boundary and at the black hole horizon, the next step is to determine a few leading terms in series expansions of $\phi(z)$ and $\psi(z)$ near $z=z_H$ and $z=0$. Assuming that $\phi(z)$ and $\psi(z)$ are smooth and infinitely differentiable functions, two asymptotes and their first derivatives may be sewed at an intermediate point on $[0,z_H]$, say at $z=1/2 z_H$. Then we will be able to determine the explicit form of the boundary operator dual to the scalar field, and to check whether it really undergoes a phase transition.

Consider series expansions of fields near $z=z_H$. We get
\[
\phi(z)=\phi(z_H)+\phi^\pr(z_H)(z-z_H)+\fr12 \phi^{\pr \pr}(z_H)(z-z_H)^2+\dots, 
\]
\[
\psi(z)=\psi(z_H)+\psi^\pr(z_H)(z-z_H)+\fr12 \psi^{\pr \pr}(z_H)(z-z_H)^2+\dots
\]
and taking into account \rf{BCzH} it becomes
\bea
\phi(z)=\phi^\pr(z_H)(z-z_H)+\fr12 \phi^{\pr \pr}(z_H)(z-z_H)^2+\dots, \nn \\
\psi(z)=\psi(z_H)+\fr12 \psi(z_H)(\fr{z}{z_H}-1)+\fr12 \psi^{\pr \pr}(z_H)(z-z_H)^2+\dots
\label{zHexp}
\eea
Since $z$ varies from $z_H$ to zero, $(z-z_H)$ is negative. $\phi(z)$ is a monotonic function which starts from zero at $z=z_H$ and takes negative values at the boundary. Therefore, $\phi^\pr(z_H)>0$. 

To fix the next coefficient in the expansion of $\phi(z)$ we use its equation of motion \rf{EOMzprobe} near the point $z=z_H$
\bea
\phi^{\pr \pr}\vert_{z=z_H}=\left[-\fr{\phi^\pr}{z}+\fr{L^2}{z^2}\fr{2\psi^2}{f}\phi \right]\vert_{z=z_H}
\label{eomphizH}
\eea 
Substituting \rf{zHexp} into \rf{eomphizH} results in
\bea
\phi(z)=\phi^\pr (z_H)(z-z_H)-\fr1{2z_H}\phi^\pr(z_H)\left(1+L^2\psi^2(z_H)\right)(z-z_H)^2+\dots
\label{expphizH}
\eea
The same procedure for $\psi(z)$ leads to
\bea
\psi(z)=\psi(z_H)+\fr12 \fr{\psi(z_H)}{z_H}(z-z_H)+\fr1{16z_H^2}\left(3-z_H^4(\phi^\pr(z_H))^2\right)\psi(z_H)(z-z_H)^2+\dots
\label{exppsizH}
\eea

The next step in realizing our program is to determine the leading coefficients in series expansions of $\phi(z)$ and $\psi(z)$ at the boundary. Having in mind the BCs \rf{BCz0} and eqs. of motion \rf{EOMzprobe}, we arrive at
\bea
\phi(z)=\m \ln z-\rho+\dots,\nn\\
\psi(z)=\psi^\2 z+\dots
\label{expz0}
\eea

Finally, we have to sew asymptotes \rf{expphizH}, \rf{exppsizH}, \rf{expz0} and their derivatives at the point $z=1/2z_H$. It leads to the following system of equations
\bea
\fr12 \psi^\2 z_H=\psi(z_H)-\fr14 \psi(z_H)+\fr1{64}\left(3-z_H^4(\phi^\pr(z_H))^2\right)\psi(z_H),\qquad (I) \nn
\\
\psi^\2=\fr{1}{2z_H}\psi(z_H)-\fr{1}{16z_H}\left(3-z_H^4(\phi^\pr(z_H))^2\right)\psi(z_H),\qquad (II) \nn \\
\m \ln(\fr{z_H}{2})-\rho=-\fr12 \phi^\pr (z_H)z_H-\fr1{8}\phi^\pr(z_H)\left(1+L^2\psi^2(z_H)\right)z_H,\qquad (III) \nn \\
\fr{2\m}{z_H}=+\phi^\pr (z_H)+\fr12 \left(1+L^2\psi^2(z_H)\right)\phi^\pr (z_H)\qquad (IV)
\label{sew1}
\eea
Eqs. (I) and (II) of \rf{sew1} result in $z_H^4(\phi^\pr(z_H))^2=41/3$ and $\psi^\2=7\psi(z_H)/6z_H$. Hence
\bea
\phi^\pr(z_H)=+\sqrt{\fr{41}{3}}\fr1{z_H^2},
\label{phiprzH}
\eea
where the positivity of $\phi^\pr(z)$ has been taken into account.

To solve eqs. (III) and (IV) of \rf{sew1} note that the charge density $\rho$ and the chemical potential $\m$ are not independent variables, since $\rho=\m\ln{z_H}$. Such a choice is compatible with $\phi(z_H)=0$ (cf. \rf{BCzH}).
Hence, (III) and (IV) come to
\bea
\m \ln(\fr{1}{2})=-\fr58 \phi^\pr (z_H)z_H-\fr1{8}\phi^\pr(z_H)L^2\psi^2(z_H)z_H, \nn \\
\fr{\m}{2}=\fr38 \phi^\pr (z_H)z_H+\fr18 \phi^\pr (z_H) L^2\psi^2(z_H)z_H,
\eea
and we get
\bea
\phi^\pr(z_H)z_H=-4\m a,\qquad \psi^2(z_H)L^2=-\fr1{a}(1+3a)
\eea
with $a=1/2+\ln(1/2)\approx -1/ 5.1774$. From the latter expressions, taking into account the Hawking temperature $2\pi T=z^{-1}_H$, and $\phi^\pr(z_H)$ from \rf{phiprzH}, we arrive at
\bea
\psi(z_H)\approx 2.275 \left(1-T/T_c \right)^{1/2}\fr1{L},
\label{psizHTc}
\eea
where the critical temperature $T_c$ is defined by
\bea
T_c=\fr{2}{\pi \sqrt{123}}\cdot \m \approx 0.057 \cdot \m
\label{Tc1}
\eea
Therefore, the scalar operator of the boundary CFT $\langle \mathcal{O} \rangle_\psi = \psi^\2=7\psi(z_H)/6z_H$ takes the following value ($L=1$) near the phase transition point
\bea
\langle \mathcal{O} \rangle_\psi \approx 12.7 \, \sqrt{TT_c} \left(1-T/T_c \right)^{1/2} \quad \stackrel{T\rightsquigarrow T_c}{\longrightarrow} \quad \langle \mathcal{O} \rangle_\psi \approx 12.7 \, T_c \left(1-T/T_c \right)^{1/2}.
\label{Opsi1}
\eea
The numerical coefficient in front of \rf{Opsi1} and the dependence of the scalar operator on $T$ are in a good agreement with that obtained in the numerical solutions to eqs. \rf{EOMzprobe} \cite{Ren10}, \cite{Liu:2011fy}, where $\langle \mathcal{O} \rangle_\psi\approx 12.2\, T_c \left(1-T/T_c \right)^{1/2}$ was obtained. The value of $T_c/\m$ is about twice less in compare to the numerical approach \cite{Ren10}, \cite{Liu:2011fy}, where $T_c/\m \approx  0.136$. However, this discrepancy is typical in the considered here analytical approach, and it could be slightly corrected by a more accurate choice of the sewing point, see Table 1 in \cite{Chen:2011en} for the conformal dimension $\l=1$.

Achieved the first goal of the paper, let's modify the model to include magnetic field and to study the supeconductor response to such modifications.

\section{Phase transition in external magnetic field}

As for a conventional superconductor, it is naturally to expect changing the characteristics of 2D holographic superconductor in external magnetic field. The critical temperature in holographic superconductor models is sensible indeed to the magnetic field, that can be seen as numerically \cite{Nakano:2008xc}, \cite{Wen:2008pb}, as well as analytically \cite{Bellon:2010xs}, \cite{Ge:2011cw}. Setups and subsequent calculations for 3D probe holographic superconductors in external magnetic field mostly exploits the dyonic black hole solution \cite{Romans:1991nq}, \cite{Hartnoll:2007ai} to $\mathrm{AdS_4}$ Einstein-Maxwell theory. Furthemore, such a solution can be generalized to the full system of equations with backreaction of matter fields to the gravitational sector. Curiously, the same construction is not available in $\mathrm{AdS_3}$.

The electrically charged rotating black hole solution to three-dimensional AdS gravity with Maxwell field is well-known \cite{Banados:1992wn}, \cite{Banados:1992gq}. This is the famous BTZ black-hole.\footnote{Note that in \cite{Banados:1992wn}, \cite{Banados:1992gq} it was established the solution to $\mathrm{AdS_3}$ equations corresponding to a neutral spinning black hole. The charged spinning black hole solution to $\mathrm{AdS_3}$ Einstein-Maxwell system was found in \cite{Clement93} and further analyzed in \cite{Martinez99}.} 
But it turns out that the magnetically charged BTZ black hole solution \cite{Clement:1995zt}, \cite{Hirschmann:1995he}, \cite{Cataldo:1996ue}, \cite{Salgado96}, \cite{Dias:2002ps} does not well supply for our purpose. Already in first papers on this subject \cite{Clement:1995zt}, \cite{Hirschmann:1995he}, it was pointed out that the so-called magnetic solution to $\mathrm{AdS_3}$ Einstein-Maxwell equations is horizonless. It does not correspond to a black hole, but is similar to a particle (magnetic monopole) or rather to a Nielsen-Olesen \cite{Nielsen:1973cs} vertex, since the magnetic BTZ-type solution still possesses the naked singularity at the origin of the coordinate system. Other interpretations of the magnetic-type BTZ solutions may be found in \cite{Dias:2002ps} and \cite{Cataldo:2004uw}. However, independently on interpretations, the main conclusion on the absence of the horizon for the magnetic solution to D=3 AdS Einstein-Maxwell system remains unchanged. 

One may wonder do appropriate modifications of $U(1)$ gauge theory change the picture, so that one ends up with $\mathrm{AdS_3}$ dyonic (or magnetic) black hole? It turns out that any viable modifications of Maxwell theory in the bulk \cite{Clement:1995zt}, \cite{Clement96}, \cite{Mansouri99}, \cite{Dereli00}, \cite{Andrade05}
result in solutions asymptotic to extreme, hence horizonless, BTZ metric.\footnote{A positive role of the Chern-Simons term in topologically massive $\mathrm{AdS_3}$ electrodynamics consists in replacing the vector-potential logarithmic divergence at the boundary with the inverse power asymptote. As a by-product one gets finite mass and angular momentum of the solution.} Furthemore, the absence of the magnetically charged solution for a generalized power-Maxwell theory (theory with $(F_{mn}F^{mn})^n, n>1$) was recently established in \cite{Unver11}. The Dirac-Born-Infeld non-linear electrodynamics falls into the class of such theories, therefore there is not dyonic black holes in 3D AdS Einstein-Born-Infeld theory.

For reasons in the above we have to figure out another way to engage the magnetic field in a game. It is worth mentioning to this end the Barnett effect \cite{Barnett35} of magnetization of uncharged, but rotated body, with magnetization proportional to the angular velocity of the sample, and the London magnetic moment \cite{London50}, which appears upon rotating a superconductor. Therefore, rotating the superconductor one puts the condensate of the Cooper pairs in the effective magnetic field.

Studying the Barnett-London effect in context of 3D holographic superconductivity was started with \cite{Sonner:2009fk}, where it was proposed to search for feedbacks of the AdS boundary rotation on the charged scalar field condensation. Needless to say, the realization of this scenario requires to deal with spinning black holes, i.e. with $\mathrm{AdS_4}$ Kerr-Newman black holes. Early, effects of rotation in the AdS/CFT correspondence were studied in \cite{Hawking:1998kw} and \cite{Berman:1999mh}, \cite{Hawking:1999dp}.

Now let's turn to the second task of the paper to establish changes in the principle characteristics of 2D CFT near the phase transition due to rotating black hole.

\section{Rotation in BTZ coordinates}

The metric of spinning neutral BTZ black hole, in its original form \cite{Banados:1992wn}, is
\bea
ds^2=-f(r)dt^2+\fr{dr^2}{f(r)}+r^2\left(d\varphi-\fr{J}{2r^2}dt\right)^2 ,
\label{BTZJ}
\eea
where
\[
f(r)=-M+\fr{r^2}{L^2}+\fr{J^2}{4r^2}.
\]
It is convenient to write down the metric \rf{BTZJ} in new coordinates $z=L^2/r$, in which it becomes
\bea
ds^2=\fr{L^2}{z^2}\left[-f(z)dt^2+\fr{dz^2}{f(z)}+L^2 \left(d\varphi-\fr{Jz^2}{2L^4}dt\right)^2\right] .
\label{Jz}
\eea
Here 
\[
f(z)=1-\fr{Mz^2}{L^2}+\fr{J^2 z^4}{4L^6},
\]
and the metric in \rf{Jz} is obviously related to \rf{AdS3BH} with setting $J=0$ and identifying $z_H^2=L^2/M$, $x=L\varphi$. Note that the horizon(s) of the spinning black hole \rf{Jz} is(are) identified from the equation
\bea
1-\fr{Mz_\pm^2}{L^2}+\fr{J^2 z_\pm^4}{4L^6}=0  \quad \leadsto \quad
z^2_\pm=\fr{2ML^4}{J^2}\left[1\pm \left(1-\fr{J^2}{M^2L^2}\right)^{1/2}\right].
\label{HorJ}
\eea
The limit $J\rightarrow 0$ should be taken with care in $(t,\varphi,z)$ coordinates; from \rf{HorJ} it comes
\bea
z^2_\pm \stackrel{J\rightarrow 0}{\leadsto} \fr{2ML^4}{J^2}\left[1- \left(1-\fr{J^2}{2M^2L^2}+\dots \right)\right]=\fr{L^2}{M}+\mathcal{O}(J^2),
\label{limzH}
\eea
and only one of two horizons $z_\pm$ survives\footnote{The other horizon $z_+$ tends to infinity and corresponds to a naked singularity, hence it has to be discarded by the cosmic censorship principle.}: $z_-=L^2/M=z_H$ from the non-rotating BTZ metric \rf{AdS3BH}. In $(t,\varphi,r)$ coordinates two horizons are
\[
r^2_\pm=\fr{ML^2}{2}\left[1\pm \sqrt{1-\fr{J^2}{M^2L^2}}\right]
\]
and the limit $J\rightarrow 0$ is smooth. Taking the limit results in arising just one horizon again, located at $r_H=ML^2$. The second horizon turns into a naked singularity, hence discarded.

Before getting the explicit form of the equations of motion, let us briefly discuss two points. 

First, 
setting $\tilde{\varphi}=L\varphi$ one observes that a massive scalar field equation 
\bea
\fr{1}{\sqrt{-g}} \pa_m(\sqrt{-g}g^{mn}\pa_n \Psi(z,\tilde{\varphi}))-m^2\Psi(z,\tilde{\varphi})=0
\label{Psim}
\eea
admits the complete separation in terms of two functions $\psi(z)$ and $S(\tilde{\varphi})$ with a separation constant $\l$. Taking into account the only $z$ dependence of $\sqrt{-g}$, one arrives at
\bea
g^{\tilde{\varphi}\tilde{\varphi}}\pa_{\tilde{\varphi}} \pa_{\tilde{\varphi}} S(\tilde{\varphi})=-\l S(\tilde{\varphi}),\qquad g^{\tilde{\varphi}\tilde{\varphi}}=\fr{z^2}{L^2}\left(1-\fr{J^2z^4}{4L^6f(z)}\right)\nn\\
\pa_z (\sqrt{-g} g^{zz}\partial_z \psi(z))-\sqrt{-g}m^2\psi(z)=\l\psi(z) .
\label{PsimEOM}
\eea
First equation of \rf{PsimEOM} admits the solution in terms of the one-dimensional Laplacian $\pa_{\tilde{\varphi}}^2$ eigenfunctions $S(\tilde{\varphi})=e^{i\a\tilde{\varphi}}$ with eigenvalues\footnote{Eigenvalues $\a$ are in general complex and unconstrained. They should not be confused with eigenvalues of a one-dimensional Laplacian in the Fourier analysis, where to form the orthogonal basis of eigenfunctions the eigenvalues are integers. In what follows we will consider only $\a \in \mathbb{R}$.}
\[
\a^2=\fr{\l L^2}{z^2} \left(1-\fr{J^2z^4}{4L^6f(z)}\right)^{-1}.
\]
It fixes the separation constant $\l$ in the second equation of \rf{PsimEOM} for the ``radial" part of $\Psi(z,\tilde{\varphi})$.

Second, we have to fix an ansatz  for the bulk gauge and the charged scalar field. For the scalar field we choose 
\[
\Psi=\Psi(z,\tilde{\varphi}),
\]
and $\Psi$ picks up the dependence on $\tilde{\varphi}$ due to the black hole rotation. Note that ``rotational" dependence of the bulk fields has no effect in unbounded (non-compact) space-times (flat Minkowski and dS). But the black hole rotation has an impact on the boundary in AdS space, due to the frame dragging (Lense-Thirring) \cite{Lense18} effect.  

For the gauge field $A_m$ one should also take $A_m(z,\tilde{\varphi})$. However, one can say more on the form of $A_m(z,\tilde{\varphi})$: it has to be\footnote{Hope the reader will not confuse between the temporal component of the gauge field $A_t=\phi(z)$ and one of the coordinates $\tilde{\varphi}$.} $A(z,\tilde{\varphi})=\phi(z)dt+\xi(z)d\tilde{\varphi}$ that is dictated by the explicit form of the charged rotating solution to the Einstein-Maxwell AdS equations of motion \cite{Clement93}, \cite{Martinez99}. 

New component of the vector potential plays a destructive role for the scalar field condensate on the AdS boundary. This conclusion directly comes from the scalar field equation of motion
\[
D_m(\sqrt{-g}g^{mn}D_n \Psi)-\sqrt{-g}m^2\Psi=0,
\]
which in the background \rf{Jz} and with the above mentioned ansatz for the gauge field transforms into
\[
\partial_m(\sqrt{-g}g^{mn}\pa_n \Psi)+f(A,\pa)\Psi-V(\Psi)=0.
\]
Here $f(A,\pa)$ is an operator of the first degree over $\pa_m$, acting on $\Psi$, and $V(\Psi)$ is an effective potential, whose explicit form is
\bea
V(\Psi)=\sqrt{-g}\left[ m^2+A_t g^{tt} A_t+A_{\tilde{\varphi}}g^{\tilde{\varphi}\tilde{\varphi}}A_{\tilde{\varphi}}\right].
\label{Veff}
\eea
In the non-rotating case, when $A_{\tilde{\varphi}}=0$, the effective mass of the scalar field decreases that makes the condensation possible \cite{Gubser:2008px}. It happens because of $g^{tt}<0$. But in the case at hand the effective mass gets decreased smaller, due to $g^{\tilde{\varphi}\tilde{\varphi}}$, that makes the condensation hard.\footnote{In fact $g^{\tilde{\varphi}\tilde{\varphi}}>0$ only for small $J$, and far from the horizon. For sufficiently large $J$, $g^{\tilde{\varphi}\tilde{\varphi}}$ turns into negative values, but the black hole becomes very unstable in the case. Note, however, that in the limit of small $J$, studied below, some special choice of parameters may cause $g^{\tilde{\varphi}\tilde{\varphi}}<0$.} Therefore, rotation affects the phase transition, and out aim is to realize manifestations of this effect.

\section{Probe limit in a rotating 2D holographic superconductor}

As in the non-rotating case we will solve equations of motion of a charged scalar interacting with abelian gauge field
\bea\label{EOM3J}
\fr1{\sqrt{-g}}D_m(\sqrt{-g}D^m\Psi)-m^2\Psi=0,\nn \\
\pa_n(\sqrt{-g}F^{nm})+\sqrt{-g}i(\Psi\overline{D^m}\Psi^*-\Psi^*D^m \Psi)=0,
\eea
in the probe limit, i.e. in the background 
\bea
ds^2=\fr{L^2}{z^2}\left[\left(-f(z)+\fr{J^2z^4}{4L^6}\right)dt^2+\fr{dz^2}{f(z)}+d\tilde{\varphi}^2-2\fr{Jz^2}{2L^3}dtd\tilde{\varphi} \right]
\label{RotBG}
\eea
apparently related to \rf{Jz}. In \rf{EOM3J} $D\Psi\equiv (\pa_m-iA_m)\Psi$, $\overline{D^m}\Psi^*=(D\Psi)^*$, and $F_{mn}=2\pa_{[m}A_{n]}$. 

With the following ansatz 
\bea
\Psi=\Psi(z,\tilde{\varphi}),\qquad A=\phi(z)dt+\xi(z)d\tilde{\varphi}
\label{anzJ}
\eea
eqs. \rf{EOM3J} become
\bea
\pa_z(\sqrt{-g}g^{zz}\pa_z \Psi)+D_{\tilde{\varphi}}(\sqrt{-g}g^{\tilde{\varphi}\tilde{\varphi}}D_{\tilde{\varphi}} \Psi)-i\sqrt{-g}A_t g^{t\tilde{\varphi}}\pa_{\tilde{\varphi}}\Psi \nn\\
-\sqrt{-g}\left( m^2+A_t g^{tt}A_t+2A_t g^{t\tilde{\varphi}}A_{\tilde{\varphi}}\right)\Psi=0, \qquad \qquad \mathrm{(SI)} \nn\\
\partial_m(\sqrt{-g}F^{mz})+i\sqrt{-g}\left(\Psi \pa^z \Psi^*-\Psi^*\pa^z \Psi \right)=0,\qquad \qquad \mathrm{(VI)} \nn\\
\pa_m(\sqrt{-g}F^{m\tilde{\varphi}})+i\sqrt{-g}\left(\Psi (D^{\tilde{\varphi}} \Psi)^*-\Psi^*D^{\tilde{\varphi}} \Psi \right)=0,\qquad \qquad \mathrm{(VII)} \nn\\
\pa_m(\sqrt{-g}F^{mt})+i\sqrt{-g}\left(\Psi (D^{t} \Psi)^*-\Psi^*D^{t} \Psi \right)=0.\qquad\qquad  \mathrm{(VIII)}
\label{EOMcJ}
\eea

Let's take a look at the scalar equation of motion (SI). To solve this equation we separate variables by $\Psi(z,\tilde{\varphi})=\psi(z)S(\tilde{\varphi})$ and substitute it into (SI).

The ``angular" part of (SI) is
\bea
\sqrt{-g}g^{\tilde{\varphi}\tilde{\varphi}}\pa^2_{\tilde{\varphi}}S(\tilde{\varphi})-i\sqrt{-g}A_t g^{t\tilde{\varphi}}\pa_{\tilde{\varphi}}S(\tilde{\varphi})-2i\sqrt{-g}A_{\tilde{\varphi}}g^{\tilde{\varphi}\tilde{\varphi}}\pa_{\tilde{\varphi}}S(\tilde{\varphi})=-\l S(\tilde{\varphi}),
\label{angSJ}
\eea
where $\l$ is the separation constant. It is fixed by observing that $S(\tilde{\varphi})=e^{i\a \tilde{\varphi}}$ are the eigenfunctions of the corresponding to (SI) differential operator; then
\bea
\l=\fr{L\a^2}{z}\left(1-\fr{J^2z^4}{4L^6f(z)}\right)+\fr{\a Jz}{2L^2 f(z)}\phi(z)-\fr{2\a L}{z}\left(1-\fr{J^2z^4}{4L^6f(z)}\right)\xi(z),
\label{lambdaJ}
\eea
where \rf{RotBG}, \rf{anzJ} have been used. 

Fixed the value of $\l$ we will solve the ``radial" part of the scalar equation. Hovewer, before doing any further steps, let us simplify the consideration since the system of equations \rf{EOMcJ} with the ansatz \rf{anzJ} does not look simple for the analytical treatment. Setting $A_{\tilde{\varphi}}=0$ makes the problem more tractable, but one may wonder on the legality of this step. Let's make two comments to this end. First, in the probe limit fields are supposed to propagate in the fixed background, so we consider the ansatze \rf{ans1} and \rf{anzJ} as small perturbations over the background. Therefore, if we consider a slow angular velocity of the black hole horizon we deal with a small angular momentum. It means that the $\tilde{\varphi}$ dependence in $A_m$ gets only the sub-leading contribution. It can be viewed, for instance, from \rf{lambdaJ}: the leading in $J$ contribution at small $J$ comes from the temporal component of $A_m$. Second, the limit of small angular momenta is reasonable in view of instability of rotating black holes with large $J$ \cite{Hawking:1999dp}, \cite{Cardoso:2004hs} related to the super-radiant scattering effect \cite{Zeldovich71}, \cite{Bardeen72}, \cite{Starobinsky73}. Another way to understand the sub-leading nature of 
$A_{\tilde{\varphi}}$ is to notice that the rotating solution with non-zero ``angular" component of the gauge field potential can be generated from the non-rotating BH solution \cite{Clement93}. Then $A_{\tilde{\varphi}}\sim \omega A_t$, where $\omega$ is the angular velocity. Clearly, $A_{\tilde{\varphi}}\ll A_t$ when $\omega \ll 1$.

\subsection{Small angular momentum approximation}

On account of the above, we will try to solve equations of motion \rf{EOMcJ} in the following setting
\bea
\Psi=\Psi(z,\tilde{\varphi}),\qquad A=\phi(z)dt+\xi(z)d\tilde{\varphi},\qquad J\ll 1,\qquad \xi(z)\ll \phi(z).
\label{anzJs}
\eea
In practice it means that we take $A\approx\phi(z)dt$, and will mostly be interested in terms linear in $J$.

Let's turn to equations \rf{EOMcJ} and to consider the ``radial" part of the scalar equation of motion.
It is
\bea
\pa_z(\sqrt{-g}g^{zz}\pa_z \psi(z))-\sqrt{-g}A_t g^{tt} A_t\psi(z)-\sqrt{-g}m^2\psi(z)\approx \l\psi(z),
\label{radScJ}
\eea
where the separation constant $\l$ is now
\bea
\l\approx\fr{L\a^2}{z}+\fr{\a J z}{2L^2 f(z)}\phi(z).
\label{lambdaJs}
\eea
Taking into account the explicit form of the inverse metric components
\bea
g^{tt}=-\fr{z^2}{L^2f(z)},\qquad g^{t\tilde{\varphi}}=-\fr{Jz^4}{2L^5f(z)},\qquad g^{zz}=\fr{z^2}{L^2}f(z),\nn\\
g^{\tilde{\varphi}\tilde{\varphi}}=\fr{z^2}{L^2}\left(1-\fr{J^2z^4}{4L^6 f(z)}\right),
\label{ginvJs}
\eea
from \rf{radScJ} we get
\bea
\psi^{\pr \pr}+\left(\fr{f^\pr}{f}-\fr{1}{z}\right)\psi^\pr+\left(\fr{\phi^2}{f^2}-\fr{m^2L^2}{z^2 f}-\fr{1}{f}\left[\a^2+\fr{\a Jz^2}{2L^3 f}\phi \right]\right)\psi\approx0,
\label{radScJf}
\eea
which is the desired equation for $\psi(z)$.

Now consider the gauge field equation of motion in its component form, eqs. (VI)-(VIII) of \rf{EOMcJ}. From (VI) it follows that the phase $\theta(z,\tilde{\varphi})$ of the charged complex scalar field $\Psi$ is constant in $z$ direction. From the next equation (VII) one may conclude that the phase is also independent on $\tilde{\varphi}$. Hence, without loss of generality, we will consider real $\Psi$, that has silently been supposed in deriving eq. \rf{radScJf}.

The system of the component vector field equations in the background \rf{ginvJs} transforms into
\bea
\pa_z(\sqrt{-g}g^{\tilde{\varphi}\tilde{\varphi}}g^{zz}F_{zt})-2\sqrt{-g}\Psi^2 g^{\tilde{\varphi}t}A_t\approx 0,\nn\\
\pa_z(\sqrt{-g}g^{zz}g^{tt}F_{zt})-2\sqrt{-g}\Psi^2 g^{tt}A_t\approx 0,
\label{AEOMJ}
\eea
where, in accordance to \rf{angSJ}, $\Psi(z,\tilde{\varphi})=\psi(z)e^{i\a \tilde{\varphi}}$ ($\psi(z) \in \mathbb{R}$). These equations combine into the single equation of motion
\[
\pa_z\left(\sqrt{-g}g^{zz}(g^{\tilde{\varphi}t}+g^{tt})F_{zt}\right)-2\sqrt{-g}\psi^2 e^{2i\a \tilde{\varphi}}(g^{\tilde{\varphi}t}+g^{tt})A_t \approx 0,
\]
that in the considered limit is
\bea
\phi^{\pr \pr}+\fr{1}{z\left(1+\fr{Jz^2}{2L^3}\right)} \left(1+\fr{3Jz^2}{2L^3}\right)\phi^\pr-\fr{2L^2}{z^2 f}\psi^2 e^{2i\a \tilde{\varphi}}\phi\approx 0.
\label{phiEOMJ}
\eea
Therefore, the system of equations for a rotating 2D holographic superconductor in the probe limit and in the slow rotation approximation comes as follows
\bea
\psi^{\pr \pr}+\left(\fr{f^\pr}{f}-\fr{1}{z}\right)\psi^\pr+\left(\fr{\phi^2}{f^2}-\fr{m^2L^2}{z^2 f}-\fr{1}{f}\left[\a^2+\fr{\a Jz^2}{2L^3 f}\phi \right]\right)\psi\approx 0, \label{psiJzfin}\\
\phi^{\pr \pr}+\fr{1}{z\left(1+\fr{Jz^2}{2L^3}\right)} \left(1+\fr{3Jz^2}{2L^3}\right)\phi^\pr-\fr{2L^2}{z^2 f}\psi^2 e^{2i\a \tilde{\varphi}}\phi\approx 0.
\label{AJzfin}
\eea
Clearly, \rf{EOMzprobe} and \rf{psiJzfin}, \rf{AJzfin} are compatible to each other once the rotation stops, i.e. under $\a \rightarrow 0$, and $J \rightarrow 0$.

\subsection{Analytic solution to the small $J$ approximation at the BF bound}

As in the non-rotating case we need to specify boundary conditions (BCs) to calculate the critical temperature and the characteristic exponent of the phase transition. BCs at the horizon $z=z_H$ are
\bea
\phi(z_H)=0,\qquad \psi^\pr(z_H)=-\fr{1}{2z_H}\left(m^2L^2+\a^2 z^2_H\right)\psi(z_H),
\label{BCzHJ}
\eea
and reasons for this choice are settled back to the non-rotating case. Near the boundary we have
\bea
\phi=\phi^\1\ln z+\phi^\2,\qquad \psi=\psi^\1 z\ln z+\psi^\2 z,
\label{BCz0J}
\eea
with some integration constants $\phi^{(i)}, \psi^{(j)}$ ($i,j=1,2$). In what follows we will fix the mass of the scalar field to be $m^2L^2=-1$ that corresponds to the BF bound.

As in the non-rotating case, we are going to define a few leading terms in the series expansions of $\phi(z)$ and $\psi(z)$ near the horizon $z=z_H$, and near the boundary $z \rightarrow 0$. Note that within the approximation $f(z)$ is defined by the same relation as in the non-rotating case, i.e.
\[
f(z)\approx 1-\fr{z^2}{z^2_H} .
\] 
Therefore, the black hole temperature does not sufficiently change, and $T\approx1/2\pi z_H$.

Near the horizon
\[
\phi(z)=\phi(z_H)+\phi^\pr (z_H)(z-z_H)+\fr{1}{2}\phi^{\pr \pr}(z_H)(z-z_H)^2+\dots,
\]
\[
\psi(z)e^{i\a \tilde{\varphi}}=\psi(z_H)+\psi^\pr (z_H)(z-z_H)+\fr{1}{2}\psi^{\pr \pr}(z_H)(z-z_H)^2+\dots,
\]
and BCs \rf{BCzHJ} have to be taken into account. 

Let's get started with evaluating a few leading coefficients in the scalar field series expansion. It is convenient to parameterize the difference between $z$ and $z_H$ as follows \cite{Balasubramanian:2010ys}
\bea
z^2=z^2_H(1-\epsilon \bar{z}^2) \qquad \leadsto \qquad z_H-z=\fr{z_H}{2}\epsilon \bar{z}^2.
\label{zHeps}
\eea
Then, the near-horizon limit corresponds to taking the $\epsilon \rightarrow 0$ limit in the end of calculations.

The leading coefficient in $\psi(z)e^{i\a \tilde{\varphi}}$ series expansion had fixed, the next coefficient is determined by the BCs \rf{BCzHJ}
\bea
\psi^\pr(z_H)=\fr{1}{2z_H}\left(1-\a^2 z^2_H\right)\psi(z_H),
\label{psipr}
\eea
and the equation of motion of $\psi$ can be used to define $\psi^{\pr \pr}(z_H)$. 

Examining $\psi^{\pr \pr}(z_H)$ one may notice that all but one terms are regular in the near-horizon limit $\epsilon \rightarrow 0$. The divergent part of $\psi^{\pr \pr}(z_H)$ follows from the last term in $\psi$ equation of motion
\[
-\fr{\a Jz^2}{2L^3 f^2}\phi(z) \psi(z)
\]
that leads to
\bea
\psi^{\pr \pr}\vert_{\epsilon \rightarrow 0} \sim \fr{\a J z^4_H}{8L^3}\left[ \phi^\pr (z_H)\psi^\pr (z_H)+\fr{1}{2}\phi^{\pr \pr}(z_H)\psi(z_H)\right]-\fr{\a J z^3_H}{4L^3\epsilon \bar{z}^2}\phi^\pr(z_H)\psi(z_H).
\label{DivJs}
\eea
The last term in \rf{DivJs} is divergent, and requires a regularization. Introducing the regularized angular momentum
\bea
J=\epsilon \bar{z}^2 \mathcal{J}_R,
\label{JReg}
\eea
in the near-horizon limit we get
\bea
\psi^{\pr \pr}\vert_{\epsilon \rightarrow 0} \sim -\fr{\a \mathcal{J}_R z^3_H}{4L^3}\phi^\pr(z_H)\psi(z_H).
\label{DivJRs}
\eea

The regular term of \rf{DivJs} disappeared from the final expression for $\psi^{\pr \pr}(z_H)$. It seems quite unnatural, but it may be explained once we figure out an hierarchy of terms in \rf{DivJs}. In the near horizon series expansions of $\phi$ and $\psi$ the leading term correspond to $(\epsilon \bar{z}^2)^0$, next terms are of the order $(\epsilon \bar{z}^2)^1$, and so on. Then, the term in \rf{DivJRs} is proportional to $\phi^\pr(z_H)\psi(z_H)$, hence it is of the order $(\epsilon \bar{z}^2)^1$, whilst the vanishing in the angular momentum regularization $[ \phi^\pr (z_H)\psi^\pr (z_H)+\fr{1}{2}\phi^{\pr \pr}(z_H)\psi(z_H) ]$ is of the order $(\epsilon \bar{z}^2)^2$, and can be neglected as a sub-leading term in the approximation.

Therefore, from $\psi$ equation of motion \rf{psiJzfin} we get
\bea
\psi^{\pr \pr}(z_H)=\fr{3}{4 z_H}\psi^{\pr}(z_H)-\fr{z_H^2}{8}(\phi^\pr (z_H))^2 \psi(z_H) \nn \\
-\fr{\a^2 z_H}{4}\psi^\pr (z_H)-\fr{\a \mathcal{J}_R z^3_H}{8L^3}\phi^\pr(z_H)\psi(z_H) ,
\label{psi"JR}
\eea
and
$$
\psi(z)=\psi(z_H)+\fr{1}{2z_H}(1-\a^2 z^2)\psi(z_H)(z-z_H)
+\fr{1}{16 z^2_H}\left[ 3(1-\a^2 z^2_H)(1-\fr{2}{3}\a^2 z^2_H) \right.
$$
\bea
\left.
-z^4_H \left((\phi^\pr)^2+\fr{\a \mathcal{J}_R z_H}{L^3} \phi^\pr (z_H)\right) \right] \psi(z_H) (z-z_H)^2+\dots
\label{psiJRz}
\eea

Now let's turn to $\phi$ equation of motion \rf{AJzfin}. Within the approximation this equation is equivalent to the following one
\bea
\phi^{\pr \pr}+\fr{1}{z}\left( 1+\fr{Jz^3}{L^3}\right) \phi^\pr-\fr{L^2}{z^2}\fr{{2\psi^2}e^{2i\a \tilde{\varphi}}}{f}\phi \approx 0.
\label{phiEOMJz}
\eea
To determine the second order, with respect to $\epsilon \bar{z}^2$, coefficient in $\phi(z)$ series expansion near the horizon we will use \rf{phiEOMJz} with regularized angular momentum $\mathcal{J}_R$. After taking the limit $\epsilon \rightarrow 0$ the equation becomes
\bea
\phi^{\pr \pr}\vert_{z=z_H}\approx \left[-\fr{\phi^\pr}{z}+\fr{L^2}{z^2}\fr{2\psi^2}{f}\phi \right]\vert_{z=z_H},
\label{eomphiJRs}
\eea 
and formally coincides with the corresponding equation in the non-rotating case. Therefore,
\bea
\phi(z)=\phi^\pr (z_H)(z-z_H)-\fr1{2z_H}\phi^\pr(z_H)\left(1+L^2\psi^2(z_H)\right)(z-z_H)^2+\dots
\label{expphizHJs}
\eea

At the boundary $\psi$ and $\phi$ behave like\footnote{As in the non-rotating case we should take into account the BCs at $z=0$ \rf{BCz0J}, and to choose $\psi^\1=0$ to deal with the true boundary CFT \cite{Marolf:2006nd} in the whole range of masses $m \ge m_{BF}$.}  
\bea
\phi(z)=\mu \ln z-\rho+\dots, \nn \\
\psi(z)e^{2i\a \tilde{\varphi}}=\psi^\2 z+\dots
\label{BCz0Js}
\eea

After establishing all the necessary ingredients (eqs. \rf{psiJRz}, \rf{expphizHJs}, \rf{BCz0Js}), we are ready to sew functions and their derivatives at an intermediate point. We choose $z=1/2 z_H$, and it leads to the following system of equations (cf. \rf{sew1})
\bea
\mathrm{(I)} \hspace{5.7cm} \fr12 \psi^\2 z_H=\psi(z_H)-\fr14 (1-\a^2 z^2_H)\psi(z_H) \hspace{1.7cm} \nn\\
+\fr1{64}\left[3(1-\a^2 z^2_H)(1-\fr23 \a^2 z^2_H)-z_H^4\left((\phi^\pr(z_H))^2+\fr{\a \mathcal{J}_R z_H}{L^3}\phi^\pr (z_H)\right) \right]\psi(z_H), \hspace{1.6cm}\nn
\\
\mathrm{(II)} \hspace{7.5cm} \psi^\2=\fr{1}{2z_H}(1-\a^2 z^2_H)\psi(z_H) \hspace{1.6cm} \nn \\
-\fr{1}{16z_H}
\left[3(1-\a^2 z^2_H)(1-\fr23 \a^2 z^2_H)-z_H^4\left((\phi^\pr(z_H))^2+\fr{\a \mathcal{J}_R z_H}{L^3}\phi^\pr (z_H)\right) \right]\psi(z_H), \hspace{1.6cm}
 \nn \\
\mathrm{(III)} \hspace{2.2cm} \m \ln(\fr{z_H}{2})-\rho=-\fr12 \phi^\pr (z_H)z_H-\fr1{8}\phi^\pr(z_H)\left(1+L^2\psi^2(z_H)\right)z_H,\hspace{1.6cm} \nn 
\eea
\bea
\mathrm{(IV)} \hspace{3.9cm} \fr{2\m}{z_H}=+\phi^\pr (z_H)+\fr12 \left(1+L^2\psi^2(z_H)\right)\phi^\pr (z_H).\hspace{2.5cm}
\label{sewJRz}
\eea
Eqs. (I) and (II) of \rf{sewJRz} result in 
\bea
\phi^\pr(z_H)=+\sqrt{\fr{41}{3}}\fr1{z_H^2}\sqrt{1+\fr{1}{41}\left[17\a^2 z^2_H+6\a^4 z^4_H+3\left(\fr{\a \mathcal{J}_R z^3_H}{2L^3}\right)^2 \right]}-\fr{\a \mathcal{J}_R z_H}{2L^3},
\label{phiprzHJRz}
\eea
where the positivity of $\phi^\pr(z)$ has been taken into account. From (III) and (IV) of \rf{sewJRz} we get
\bea
\phi^\pr(z_H)z_H=-4\m a,\qquad \psi^2(z_H)L^2=-\fr{1}{a}\left(1+3a \right).
\label{psiJRza}
\eea
Here $a=1/2+\ln(1/2)$, and we have expressed the charged density $\rho$ through the chemical potential $\m$.

To evaluate the critical temperature dependence on $\mathcal{J}_R$ and the boundary scalar operator expectation value, we make the series expansion over a small $\a$ and keep the terms linear in $\mathcal{J}_R$. It leads to
\bea
\phi^\pr(z_H)\approx \fr{1}{z^2_H}\sqrt{\fr{41}3}\left(1+\fr{17}{82}\a^2 z^2_H-\fr{\a \mathcal{J}_R z^3_H}{2L^3}\right),
\label{phiJRzapp}
\eea
and using \rf{psiJRza} one arrives at
\bea
\psi^2(z_H)L^2\approx -\fr{1}{a}\left(1-\fr{\sqrt{123}}{4\m}\left[2\pi T+\fr{17\a^2}{164\pi T}-\fr{\a \mathcal{J}_R}{8\pi^2T^2L^3}\right]\right).
\label{psiJRzapp}
\eea

The critical temperature at which the phase transition occurs is defined by the relation
\bea
1-\fr{\sqrt{123}}{4\m}\left[2\pi T_c+\fr{17\a^2}{164\pi T_c}-\fr{\a \mathcal{J}_R}{8\pi^2T_c^2L^3}\right]=0.
\label{TcJRz}
\eea
This is a third order algebraic equation over $T_c$, that can formally be resolved by use of the Cardano's formulae. However, the so obtained result is not suitable to analyze the dependence of $T_c$ on $\mathcal{J}_R$, and to write down \rf{psiJRzapp} in a form close to eq. \rf{psizHTc}. It is more instructive to calculate the expectation value of the boundary scalar field operator $\langle \mathcal{O} \rangle_{\psi}=7\psi(z_H)/6z_H$
\bea
\langle \mathcal{O} \rangle_{\psi} \approx 16.68 \, T \left(1-\fr{\sqrt{123}}{4\m}\left[2\pi T+\fr{17\a^2}{164\pi T}-\fr{\a \mathcal{J}_R}{8\pi^2T^2L^3}\right] \right)^{1/2},
\label{OpsiJRz}
\eea
and to put the temperature dependence of $\langle \mathcal{O} \rangle_{\psi}/T$ ($L=1$, $\m=1$) on the plot, with different sets of parameters $(\a,\mathcal{J}_R)$, see Fig. \rf{figure:fig1}.
\begin{figure}[h]
\centering
\includegraphics[width=0.5\textwidth]{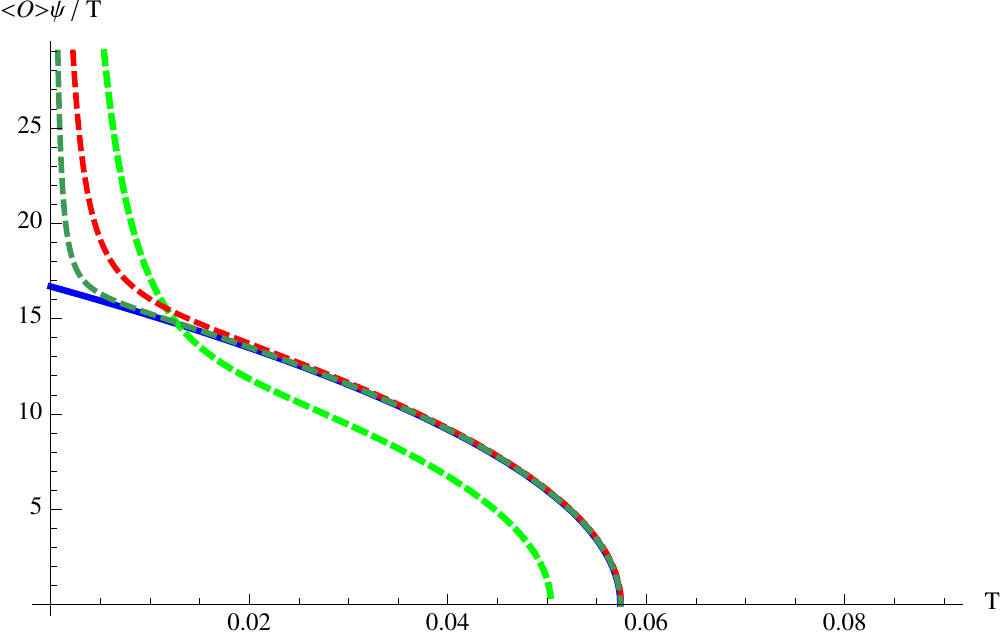}
\caption{\small $\langle \mathcal{O} \rangle_{\psi}(T)/T$ for different values of $\a$ ($\mathcal{J}_R$ is fixed to $0.0$ and $0.01$; $L=1, \m=1$): 
$(0.0,0.0)$ (blue curve), $(0.003,0.01)$ (deep green), $(0.03,0.01)$ (red), $(0.3,0.01)$ (green). } \label{figure:fig1}
\end{figure}

Looking at Fig. \rf{figure:fig1} one may notice that at fixed small $\mathcal{J}_R=0.01$ the smaller the value $\a$, the resulting curve is closer to the non-rotating dependence of $\langle \mathcal{O} \rangle_{\psi}/T$ (blue curve). Once $\a$ becomes large enough ($\a > 0.08$ as it follows from numerical simulations), the critical temperature becomes to decrease, making the condensation hard. However, the opposite case, when the critical temperature of the phase transition becomes slightly higher in compare to the non-rotating case, can be realized as well, see Fig. \rf{figure:fig2}.
\begin{figure}[h]
\centering
\includegraphics[width=0.5\textwidth]{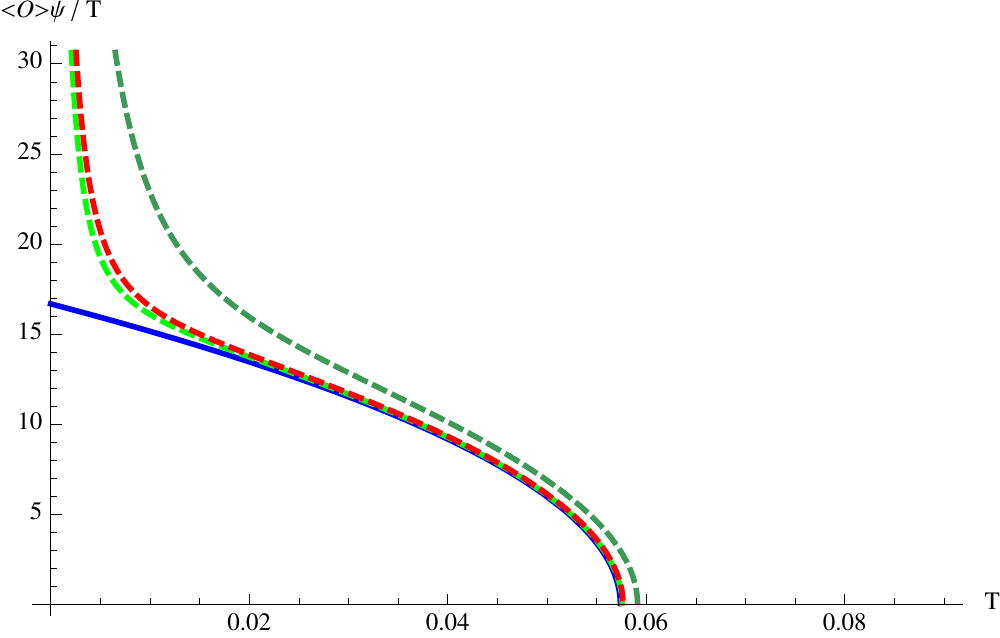}
\caption{\small $\langle \mathcal{O} \rangle_{\psi}(T)/T$ for different values of $\mathcal{J}_R$ ($\a$ is fixed to $0.003$; $L=1, \m=1$): 
$(0.0,0.0)$ (blue curve), $(0.003,0.01)$ (green), $(0.003,0.15)$ (red), $(0.003,1)$ (deep green).} \label{figure:fig2}
\end{figure}

\section{Summary and concluding remarks}

To summarize, we have filled the gap in applying the analytical methods to a 2D holographic superconductor, within the original setup of \cite{Hartnoll:2008vx}, and in the probe limit. The estimated value of the critical temperature and the temperature dependence of the boundary scalar operator have been compared with that obtained in numerical studies \cite{Ren10}, \cite{Liu:2011fy}. We have found a good agreement between the boundary scalar operator expectation value in two approaches, but the value of the critical temperature, estimated analytically, is about twice less than that reproduced in the numerical calculations. However, this discrepancy is typical within the approach we followed, as it comes from Table 1 in \cite{Chen:2011en}.

We have also extended a simple 2D holographic superconductor model to the background of a rotating BTZ black hole. It turns out that this is the way to put the superconductor into external magnetic field, since the naive extension of a charged non-rotating BTZ black hole to the magnetic $\mathrm{AdS_3}$ BH solution does not lead to the desired properties of the background. All the known magnetically charged BTZ black holes are horizonless, hence their Hawking temperature is equal to zero, that does not serve well for our purposes. However, as in the conventional superconductivity, the Barnett effect of magnetization of a rotated body \cite{Barnett35} leads to arising the London moment \cite{London50}, so one may model a weak external magnetic field by a slow superconductor rotating. In the considered model magnetization occurs once the $\mathrm{AdS_3}$ boundary gets rotated, due to the Lense-Thirring \cite{Lense18} dragging force effect. Occurrence of magnetization in a rotating holographic superconductor may be viewed in a formal extension of the ansatz \rf{ans1} with the magnetic component of the vector potential $A_z=B(z)$
\bea
\Psi=\psi(z),\qquad A=\phi(z)dt+B(z) dz.
\label{ans2}
\eea
Then eqs. \rf{EOMzprobe} are modified to 
\bea
\psi^{\pr\pr}+\left(\fr{f^\pr}{f}-\fr1{z}\right)\psi^\pr+\left(\fr{\phi^2}{f^2}-\fr{m^2L^2}{z^2 f}\right)\psi-\fr1{f}B^2 \psi=0,\nn\\
\phi^{\pr\pr}+\fr1{z}\phi^\pr-\fr{L^2}{z^2}\fr{2\psi^2}{f}\phi=0,\nn\\
B^{\pr\pr}+\left(\fr{f^\pr}{f}+\fr1{z}\right)B^\pr+\fr{L^2}{z^2}\fr{2\psi^2}{f}B=0.
\label{EOMzprobe2}
\eea
Comparing the first equation of \rf{EOMzprobe2} with the corresponding equation \rf{psiJzfin}, one observes that the external magnetic field $B$ is modeled by $\a^2+(\a J z^2/2L^3 f)\phi$, which depends on two free parameters of the problem.

We have analytically calculated, in the probe limit and in the small angular momentum approximation, the temperature dependence of the boundary scalar operator expectation value in the model of a rotating 2D holographic superconductor. Putting the obtained curves on the plot, we have observed that, in dependence on the choice of parameters $(\a,\mathcal{J}_R)$, the critical temperature of the phase transition decreases, making the condensation hard, or slightly increase in compare to the non-rotating case. It could be interesting to observe the latter effect in the numerical studies of the model, in its full complexity, and out of the small angular momentum approximation.

Further development of the model beyond the scope of the present paper consists in its extension to the full-fledged case with backreaction of fields in the bulk, and to the estimation of the DC conductivity. It would also be instructive to reformulate the model in the Boyer-Lindquist coordinates (see \cite{Krishnan09}, \cite{Krishnan10} for BTZ black hole), and to make the problem closer to a 3D rotating holographic superconductor \cite{Sonner:2009fk}.
We hope to return to these problems in subsequent publications.

\subsection*{Acknowledgments.} Illuminating discussions with V.P. Berezovoj and Yu.L. Bolotin are kindly acknowledged. Work is supported in part by the Joint DFFD-RFBR Grant \# F40.2/040.


\begin{thebibliography}{20}



\bibitem{Bardeen:1957mv}
  J.~Bardeen, L.~N.~Cooper and J.~R.~Schrieffer,
  ``Theory of superconductivity,''
  Phys.\ Rev.\  {\bf 108}, 1175 (1957).

\bibitem{Bogoliubov:1958}
N.~N.~Bogoliubov, ``A new method in the theory of superconductivity 1.," Sov. Phys. JETP, {\bf 7}, 41–46, (1958).

\bibitem{Bednorz:1986tc}
  J.~G.~Bednorz and K.~A.~Muller,
  ``Possible high Tc superconductivity in the Ba-La-Cu-O system,''
  Z.\ Phys.\  B {\bf 64}, 189 (1986).



\bibitem{AndersonBook}
P.~W.~Anderson, The Theory of Superconductivity in High-$\mathrm{T_c}$ Cuprates, Princeton University Press, 1997.

\bibitem{SachdevBook}
S.~Sachdev, Quantum Phase Transitions, CUP, 1998.


\bibitem{Chen:2008}
T.~Y.~Chen, Z.~Tesanovic, R.~H.~Liu, X.~H.~Chen and C.~L.~Chien, ``BCS-like gap in the superconductor $\mathrm{SmFeAsO_{0.85}F_{0.15}}$," Nature {\bf 453}, 1224 (2008).


\bibitem{Maldacena:1997re}
  J.~M.~Maldacena,
  ``The Large N limit of superconformal field theories and supergravity,''
  Adv.\ Theor.\ Math.\ Phys.\  {\bf 2}, 231 (1998)
  [Int.\ J.\ Theor.\ Phys.\  {\bf 38}, 1113 (1999)]
  [arXiv:hep-th/9711200].

\bibitem{Gubser:1998bc}
  S.~S.~Gubser, I.~R.~Klebanov and A.~M.~Polyakov,
  ``Gauge theory correlators from noncritical string theory,''
  Phys.\ Lett.\  B {\bf 428}, 105 (1998)
  [arXiv:hep-th/9802109].

\bibitem{Witten:1998qj}
  E.~Witten,
  ``Anti-de Sitter space and holography,''
  Adv.\ Theor.\ Math.\ Phys.\  {\bf 2}, 253 (1998)
  [arXiv:hep-th/9802150].

\bibitem{Aharony:1999ti}
  O.~Aharony, S.~S.~Gubser, J.~M.~Maldacena, H.~Ooguri and Y.~Oz,
  ``Large N field theories, string theory and gravity,''
  Phys.\ Rept.\  {\bf 323}, 183 (2000)
  [arXiv:hep-th/9905111].


\bibitem{Matveev:1973ra}
  V.~A.~Matveev, R.~M.~Muradian and A.~N.~Tavkhelidze,
  ``Automodellism in the large - angle elastic scattering and structure of
  hadrons,''
  Lett.\ Nuovo Cim.\  {\bf 7}, 719 (1973).


\bibitem{Brodsky:1973kr}
  S.~J.~Brodsky and G.~R.~Farrar,
  ``Scaling Laws at Large Transverse Momentum,''
  Phys.\ Rev.\ Lett.\  {\bf 31}, 1153 (1973).


\bibitem{Polchinski:2001tt}
  J.~Polchinski and M.~J.~Strassler,
  ``Hard scattering and gauge / string duality,''
  Phys.\ Rev.\ Lett.\  {\bf 88}, 031601 (2002)
  [arXiv:hep-th/0109174].


\bibitem{Brower:2002er}
  R.~C.~Brower and C.~I.~Tan,
  ``Hard scattering in the M theory dual for the QCD string,''
  Nucl.\ Phys.\  B {\bf 662}, 393 (2003)
  [arXiv:hep-th/0207144].



\bibitem{BoschiFilho:2002zs}
  H.~Boschi-Filho and N.~R.~F.~Braga,
  ``QCD / string holographic mapping and high-energy scattering amplitudes,''
  Phys.\ Lett.\  B {\bf 560}, 232 (2003)
  [arXiv:hep-th/0207071].


\bibitem{Andreev:2002aw}
  O.~Andreev,
  ``Scaling laws in hadronic processes and string theory,''
  Phys.\ Rev.\  D {\bf 67}, 046001 (2003)
  [arXiv:hep-th/0209256].




\bibitem{Hartnoll:2008vx}
  S.~A.~Hartnoll, C.~P.~Herzog and G.~T.~Horowitz,
  ``Building a Holographic Superconductor,''
  Phys.\ Rev.\ Lett.\  {\bf 101}, 031601 (2008)
  [arXiv:0803.3295 [hep-th]].

\bibitem{Hartnoll:2008kx}
  S.~A.~Hartnoll, C.~P.~Herzog and G.~T.~Horowitz,
  ``Holographic Superconductors,''
  JHEP {\bf 0812}, 015 (2008)
  [arXiv:0810.1563 [hep-th]].


\bibitem{GGdualBook11}
From Gravity to Thermal Gauge Theories: The AdS/CFT Correspondence (Lecture Notes in Physics),
E. ~Papantonopoulos ed., Springer, 2011.


\bibitem{Hartnoll:2009sz}
  S.~A.~Hartnoll,
  ``Lectures on holographic methods for condensed matter physics,''
  Class.\ Quant.\ Grav.\  {\bf 26}, 224002 (2009)
  [arXiv:0903.3246 [hep-th]].


\bibitem{Herzog:2009xv}
  C.~P.~Herzog,
  ``Lectures on Holographic Superfluidity and Superconductivity,''
  J.\ Phys.\ A  {\bf 42}, 343001 (2009)
  [arXiv:0904.1975 [hep-th]].



\bibitem{Horowitz:2010gk}
  G.~T.~Horowitz,
  ``Introduction to Holographic Superconductors,''
  arXiv:1002.1722 [hep-th].


\bibitem{Hertog:2006rr}
  T.~Hertog,
  ``Towards a Novel no-hair Theorem for Black Holes,''
  Phys.\ Rev.\  D {\bf 74}, 084008 (2006)
  [arXiv:gr-qc/0608075].


\bibitem{Gubser:2008px}
  S.~S.~Gubser,
  ``Breaking an Abelian gauge symmetry near a black hole horizon,''
  Phys.\ Rev.\  D {\bf 78}, 065034 (2008)
  [arXiv:0801.2977 [hep-th]].


\bibitem{Breitenlohner:1982bm}
  P.~Breitenlohner and D.~Z.~Freedman,
  ``Positive Energy in anti-De Sitter Backgrounds and Gauged Extended
  Supergravity,''
  Phys.\ Lett.\  B {\bf 115}, 197 (1982).


\bibitem{Gregory09}
  R.~Gregory, S.~Kanno and J.~Soda,
  ``Holographic Superconductors with Higher Curvature Corrections,''
  JHEP {\bf 0910}, 010 (2009)
  [arXiv:0907.3203 [hep-th]].

\bibitem{Herzog:2010vz}
  C.~P.~Herzog,
  ``An Analytic Holographic Superconductor,''
  Phys.\ Rev.\  D {\bf 81}, 126009 (2010)
  [arXiv:1003.3278 [hep-th]].

\bibitem{Bellon:2010xs}
  M.~Bellon, E.~F.~Moreno and F.~A.~Schaposnik,
  ``A Note on Holography and Phase Transitions,''
  arXiv:1012.4496 [hep-th].


\bibitem{Siopsis:2010uq}
  G.~Siopsis and J.~Therrien,
  ``Analytic calculation of properties of holographic superconductors,''
  JHEP {\bf 1005}, 013 (2010)
  [arXiv:1003.4275 [hep-th]].


\bibitem{Siopsis:2010pi}
  G.~Siopsis, J.~Therrien and S.~Musiri,
  ``Holographic conductors near the Breitenlohner-Freedman bound,''
  arXiv:1011.2938 [hep-th].

\bibitem{Chen:2011en}
  C.~M.~Chen and M.~F.~Wu,
  ``An Analytic Analysis of Phase Transitions in Holographic Superconductors,''
  arXiv:1103.5130 [hep-th].

\bibitem{Ge:2011cw}
  X.~H.~Ge,
  ``Analytical calculation on critical magnetic field in holographic
  superconductors with backreaction,''
  arXiv:1105.4333 [hep-th].



\bibitem{Carlip:2005zn}
  S.~Carlip,
  ``Conformal field theory, (2+1)-dimensional gravity, and the BTZ black
  hole,''
  Class.\ Quant.\ Grav.\  {\bf 22}, R85 (2005)
  [arXiv:gr-qc/0503022].


\bibitem{Arutuynov:2008}
K.~Yu.~Arutyunov, D.~S.~Golubev and A.~D.~Zaikin, ``Superconductivity in one dimension," Phys.\ Rep.\ {\bf 464} 1-70 (2008) [arXiv:0805.2118v4 [cond-mat.supr-con]].


\bibitem{Ren10}
J.~Ren, ``One-dimensional holographic superconductor from $AdS_3/CFT_2$
  correspondence,''
  JHEP {\bf 1011}, 055 (2010)
  [arXiv:1008.3904 [hep-th]].	


\bibitem{Liu:2011fy}
  Y.~Liu, Q.~Pan and B.~Wang,
  ``Holographic superconductor developed in BTZ black hole background with
  backreactions,''
  arXiv:1106.4353 [hep-th].


\bibitem{Deser:1982vy}
  S.~Deser, R.~Jackiw and S.~Templeton,
  ``Three-Dimensional Massive Gauge Theories,''
  Phys.\ Rev.\ Lett.\  {\bf 48}, 975 (1982).


\bibitem{Deser:1981wh}
  S.~Deser, R.~Jackiw and S.~Templeton,
  ``Topologically Massive Gauge Theories,''
  Annals Phys.\  {\bf 140}, 372 (1982)
  [Erratum-ibid.\  {\bf 185}, 406 (1988)]
  [Annals Phys.\  {\bf 185}, 406 (1988)]
  [Annals Phys.\  {\bf 281}, 409 (2000)].

\bibitem{Lashkari:2010ak}
  N.~Lashkari,
  ``Holographic Symmetry-Breaking Phases in AdS$_3$/CFT$_2$,''
  arXiv:1011.3520 [hep-th].


\bibitem{Sonner:2009fk}
  J.~Sonner,
  ``A Rotating Holographic Superconductor,''
  Phys.\ Rev.\  D {\bf 80}, 084031 (2009)
  [arXiv:0903.0627 [hep-th]].



\bibitem{Marolf:2006nd}
  D.~Marolf and S.~F.~Ross,
  ``Boundary Conditions and New Dualities: Vector Fields in AdS/CFT,''
  JHEP {\bf 0611}, 085 (2006)
  [arXiv:hep-th/0606113].

\bibitem{Hartnoll:2007ai}
  S.~A.~Hartnoll and P.~Kovtun,
  ``Hall conductivity from dyonic black holes,''
  Phys.\ Rev.\  D {\bf 76}, 066001 (2007)
  [arXiv:0704.1160 [hep-th]].




\bibitem{Nakano:2008xc}
  E.~Nakano and W.~Y.~Wen,
  ``Critical magnetic field in a holographic superconductor,''
  Phys.\ Rev.\  D {\bf 78}, 046004 (2008)
  [arXiv:0804.3180 [hep-th]].

\bibitem{Wen:2008pb}
  W.~Y.~Wen,
  ``Inhomogeneous magnetic field in AdS/CFT superconductor,''
  arXiv:0805.1550 [hep-th].


\bibitem{Romans:1991nq}
  L.~J.~Romans,
  ``Supersymmetric, cold and lukewarm black holes in cosmological
  Einstein-Maxwell theory,''
  Nucl.\ Phys.\  B {\bf 383}, 395 (1992)
  [arXiv:hep-th/9203018].


\bibitem{Banados:1992wn}
  M.~Banados, C.~Teitelboim and J.~Zanelli,
  ``The Black hole in three-dimensional space-time,''
  Phys.\ Rev.\ Lett.\  {\bf 69}, 1849 (1992)
  [arXiv:hep-th/9204099].

\bibitem{Banados:1992gq}
  M.~Banados, M.~Henneaux, C.~Teitelboim and J.~Zanelli,
  ``Geometry of the (2+1) black hole,''
  Phys.\ Rev.\  D {\bf 48}, 1506 (1993)
  [arXiv:gr-qc/9302012].


\bibitem{Clement93}
G.~Clement, 
``Classical solutions in three-dimensional Einstein-Maxwell cosmological gravity,"
Class. Quant. Grav. {\bf 10}, L49 (1993)

\bibitem{Martinez99}
C.~Martinez, C.~Teitelboim and J.~Zanelli,
``Charged Rotating Black Hole in Three Spacetime Dimensions,"  
Phys.\ Rev.\  D {\bf 61}, 104013 (2000)
  [arXiv:hep-th/9912259].

\bibitem{Clement:1995zt}
  G.~Clement,
  ``Spinning charged BTZ black holes and selfdual particle - like solutions,''
  Phys.\ Lett.\  B {\bf 367}, 70 (1996)
  [arXiv:gr-qc/9510025].


\bibitem{Hirschmann:1995he}
  E.~W.~Hirschmann and D.~L.~Welch,
  ``Magnetic solutions to (2+1) gravity,''
  Phys.\ Rev.\  D {\bf 53}, 5579 (1996)
  [arXiv:hep-th/9510181].

\bibitem{Cataldo:1996ue}
  M.~Cataldo and P.~Salgado,
  ``Static Einstein-Maxwell solutions in (2+1)-dimensions,''
  Phys.\ Rev.\  D {\bf 54}, 2971 (1996).

\bibitem{Salgado96}
M.~Cataldo and P.~Salgado, 
``The Einstein-Maxwell Extreme BTZ Black Hole with Self (Anti-self) Dual Maxwell Field," [arXiv:gr-qc/9611004].


\bibitem{Dias:2002ps}
  O.~J.~C.~Dias and J.~P.~S.~Lemos,
  ``Rotating magnetic solution in three-dimensional Einstein gravity,''
  JHEP {\bf 0201}, 006 (2002)
  [arXiv:hep-th/0201058].





\bibitem{Cataldo:2004uw}
  M.~Cataldo, J.~Crisostomo, S.~del Campo and P.~Salgado,
  ``On magnetic solution to (2+1) Einstein-Maxwell gravity,''
  arXiv:hep-th/0401189.

\bibitem{Nielsen:1973cs}
  H.~B.~Nielsen and P.~Olesen,
  ``Vortex Line Models for Dual Strings,''
  Nucl.\ Phys.\  B {\bf 61} (1973) 45.


\bibitem{Clement96}
K.~A.~Moussa and G.~Clement,
``Topologically massive gravito-electrodynamics: exact solutions,"
arXiv:gr-qc/9602034.

\bibitem{Mansouri99}
S.~Fernando and F.~Mansouri,
``Rotating Charged Solutions to Einstein-Maxwell-Chern-Simons Theory in 2+1 Dimensions,"
arXiv:gr-qc/9705016.

\bibitem{Dereli00}
T.~Dereli and Yu.~N.~Obukhov,
``General analysis of self-dual solutions for the Einstein-Maxwell-Chern-Simons theory in (1+2) dimensions,"
Phys.\ Rev.\  D {\bf 62}, 024013 (2000)
[arXiv:gr-qc/0001017].


\bibitem{Andrade05}
T.~Andrade, M.~Banados and R.~Benguria,
``The 2+1 charged black hole in topologically massive Electrodynamics,"
arXiv:hep-th/0503095.



\bibitem{Unver11}
S.~H.~Mazharimousavi, O.~Gurtug, M.~Halilsoy and O.~Unver,
``2 + 1 dimensional magnetically charged solutions in Einstein - Power - Maxwell theory,"
arXiv:1103.5646 [gr-qc].


\bibitem{Barnett35}
S.~J.~Barnett, ``Magnetization by Rotation," Physical Review {\bf 6} (1915) 239–270;\\ 
S.~J.~Barnett, ``Gyromagnetic and Electron-Inertia Effects," Rev. Mod. Phys. {\bf 7}, (1935) 129–166.

\bibitem{London50}
F.~London, ``Superfluids: Vol. 1, Macroscopic Theory Superconductivity," Dover 1950.




\bibitem{Hawking:1998kw}
  S.~W.~Hawking, C.~J.~Hunter and M.~Taylor,
  ``Rotation and the AdS / CFT correspondence,''
  Phys.\ Rev.\  D {\bf 59}, 064005 (1999)
  [arXiv:hep-th/9811056].

\bibitem{Berman:1999mh}
  D.~S.~Berman and M.~K.~Parikh,
  ``Holography and rotating AdS black holes,''
  Phys.\ Lett.\  B {\bf 463}, 168 (1999)
  [arXiv:hep-th/9907003].

\bibitem{Hawking:1999dp}
  S.~W.~Hawking and H.~S.~Reall,
  ``Charged and rotating AdS black holes and their CFT duals,''
  Phys.\ Rev.\  D {\bf 61}, 024014 (2000)
  [arXiv:hep-th/9908109].




\bibitem{Lense18}
J.~Lense and H.~Thirring,  
``\"Uber den Einfluss der Eigenrotation der Zentralk\"orper auf die Bewegung der Planeten und Monde nach der Einsteinschen Gravitationstheorie," Physikalische Zeitschrift {\bf 19}, 156–163 (1918).





\bibitem{Cardoso:2004hs}
  V.~Cardoso and O.~J.~C.~Dias,
  ``Small Kerr-anti-de Sitter black holes are unstable,''
  Phys.\ Rev.\  D {\bf 70}, 084011 (2004)
  [arXiv:hep-th/0405006].


\bibitem{Zeldovich71} 
Ya.~B.~Zeldovich, JETP Lett. {\bf 14}, 180 (1971); Sov. Phys. JETP {\bf 35}, 1085 (1972).


\bibitem{Bardeen72}
J.~M.~Bardeen, W.~H.~Press and S.~A.~Teukolsky, Astrophys. J. {\bf 178}, 347 (1972).  


\bibitem{Starobinsky73}
A.~A.~Starobinsky, Sov. Phys. JETP {\bf 37}, 28 (1973); A.~A.~Starobinsky and S.~M.~Churilov, Sov. Phys. JETP {\bf 38}, 1 (1973).


\bibitem{Balasubramanian:2010ys}
  V.~Balasubramanian, J.~Parsons and S.~F.~Ross,
  ``States of a chiral 2d CFT,''
  Class.\ Quant.\ Grav.\  {\bf 28}, 045004 (2011)
  [arXiv:1011.1803 [hep-th]].


\bibitem{Krishnan09}
C.~Krishnan,
``Tomograms of Spinning Black Hole,"
[arXiv:0911.0597 [hep-th]].

\bibitem{Krishnan10}
C.~Krishnan,
``Black Hole Vacua and Rotation,"
[arXiv:1005.1629 [hep-th]].

\end{thebibliography}
\end{document}